\documentclass[12pt]{article}
\usepackage{a4wide}
\usepackage{amssymb}
\usepackage{amsmath}
\usepackage{graphicx}
\usepackage{mathdots}
\usepackage{slashed}
\usepackage{verbatim}
\usepackage{xcolor}
\usepackage{caption}
\usepackage{subcaption}

\usepackage{array}
\usepackage{amsthm}

 \numberwithin{equation}{section}

\begin{document}

\mbox{}
\vspace{0truecm}
\linespread{1.1}


\centerline{\LARGE \bf Holographic thermal correlators revisited}
%

\medskip

\vspace{.4cm}

 \centerline{\LARGE \bf }

\vspace{1.5truecm}

\centerline{
    { \bf Hare Krishna ${}^{a}$} \footnote{harekrishna.harekrishna@stonybrook.edu}
   {\bf and}
    { \bf D. Rodriguez-Gomez${}^{b,c}$} \footnote{d.rodriguez.gomez@uniovi.es}}

\vspace{1cm}
\medskip
\centerline{{\it ${}^a$ C. N. Yang Institute for Theoretical Physics, Stony Brook University}} \centerline{{\it Stony Brook, NY 11794, USA}}
\medskip
\centerline{{\it ${}^b$ Department of Physics, Universidad de Oviedo}} \centerline{{\it C/ Federico Garc\'ia Lorca  18, 33007  Oviedo, Spain}}
\medskip
\centerline{{\it ${}^c$  Instituto Universitario de Ciencias y Tecnolog\'ias Espaciales de Asturias (ICTEA)}}\centerline{{\it C/~de la Independencia 13, 33004 Oviedo, Spain.}}
\vspace{1cm}

\centerline{\bf ABSTRACT}
\medskip 

We study 2-point correlation functions for scalar operators in position space through holography including bulk cubic couplings as well as higher curvature couplings to the square of the Weyl tensor. We focus on scalar operators with large conformal dimensions. This allows us to use the geodesic approximation for propagators. In addition to the leading order contribution, captured by geodesics anchored at the insertion points of the operators on the boundary and probing the bulk geometry thoroughly studied in the literature, the first correction is given by a Witten diagram involving both the bulk cubic coupling and the higher curvature couplings. As a result, this correction is proportional to the VEV of a neutral operator $O_k$ and thus probes the interior of the black hole exactly as in the case studied by Grinberg and Maldacena \cite{Grinberg:2020fdj}. The form of the correction matches the general expectations in CFT and allows to identify the contributions of $T^nO_k$ (being $T^n$ the general contraction of $n$ energy-momentum tensors) to the 2-point function. This correction is actually the leading term for off-diagonal correlators (\textit{i.e.} correlators for operators of different conformal dimension), which can then be computed holographically in this way.

\noindent

\newpage

\tableofcontents

\section{Introduction and conclusions}

Thermal 2-point correlation functions have a much more intricate structure than their zero temperature counterparts. One way to argue for this  is to recall that a finite temperature lorentzian CFT in $d$ dimensions is equivalent to the euclidean theory on $S^1\times \mathbb{R}^{d-1}$. The thermal circle provides both a scale --its length $\beta$-- and a direction, which allow primary operators to take a VEV (translational invariance is still a symmetry, so descendants do not get a VEV). Then, for operators inserted within a ball of size smaller than $\beta$ (so that it does not ``wind around the thermal circle"), one can use the standard flat space OPE, where, due to the non-zero VEV's, all primaries manifest themselves as shown in \cite{Iliesiu:2018fao}. 

Very recently thermal 2-point functions have been studied for holographic CFT's in \cite{Rodriguez-Gomez:2021pfh,Rodriguez-Gomez:2021mkk}.\footnote{These are related to heavy-heavy-light-light 4-point correlators in flat space through the Eigenstate Thermalization Hypothesis, see for instance \cite{Karlsson:2021duj,Fitzpatrick:2019efk,Fitzpatrick:2019zqz}. Also \cite{Alday:2020eua} has recently considered, albeit from a slightly different point of view, holographic thermal correlators.} In principle, these are read-off from the solution to the corresponding wave equation with the usual holographic recipe. As this is a daunting task (specially in order to explore the whole thermal circle), \cite{Rodriguez-Gomez:2021pfh,Rodriguez-Gomez:2021mkk} specialized to 2-point functions of scalar operators of large conformal dimension (but still much smaller than the central charge of the theory). In that limit, the Klein-Gordon equation can be solved through the WKB approximation, which in the end becomes the (exponentiated) geodesic length. Thus the problem becomes akin to the computation of geodesics in the black brane background. In this paper we will assume this limit, so that propagators, be it bulk-to-bulk or boundary-to-bulk, will be computed by the exponentiated length of a suitable geodesic.\footnote{Let us stress that this discussion pertains the computation of propagators, which then enter whatever Witten diagram we may compute (see below). The 2-point function can also we written, formally, as an interacting Witten diagram by ``splitting it into two parts and joining them again", but in the end it has to correspond to a regular geodesic.}

The starting point is the KG equation which arises from the leading term in the effective gravitational action. For instance, in the case of 4d $\mathcal{N}=4$ SYM at $T=0$ dual to IIB string theory on $AdS_5\times S^5$, the reduction of the fluctuations of the gravity theory on the $S^5$ gives rise to an effective action on $AdS_5$ whose first terms, for a particular set of fluctuations, were computed in \cite{Lee:1998bxa}. The leading term is the free field action from which the KG follows. Higher terms represent interactions giving rise to Witten diagrams with corrections. In addition to these corrections, in principle there can be higher curvature corrections in the starting SUGRA lagrangian. These results both in corrections to the gravitational background ($\alpha'$ corrections in the $AdS_5\times S^5$ case) as well as new couplings giving rise to new Witten diagrams.

The study of geodesics in black hole backgrounds, in particular in connection with thermal correlation functions, has been thoroughly considered (\textit{e.g.} \cite{Fidkowski:2003nf,Festuccia:2005pi,Hubeny:2006yu,Hubeny:2012ry}). In particular, since geodesics can probe the interior of the black hole, 2-point functions may offer a very interesting window into black hole physics. This has been  revived recently in \cite{Grinberg:2020fdj}, albeit for 1-point functions, where it was argued how thermal VEV's can encode the (appropriately renormalized) proper time to the singularity of a radially in-falling particle  through higher curvature couplings (in particular, to $W^2$, the Weyl tensor squared). In \cite{Rodriguez-Gomez:2021pfh} it was argued that a version of that mechanism would be at play in the 2-point function, which then contains new signatures from the black hole interior. Even though bottom-up, morally speaking the set-up underlying \cite{Rodriguez-Gomez:2021pfh} is that of charged operators --for instance, CPO's in $\mathcal{N}=4$ SYM--, which suggests a particular version of the higher curvature coupling of the form $|\phi_k|^2W^2$ ($\phi_k$ stands for the bulk field dual to the boundary scalar operator of interest $O_k$), giving rise to a Witten diagram which encodes in a similar manner to \cite{Grinberg:2020fdj} the time of travel to the singularity (as well as exhibits other interesting features such as the quasinormal frequencies similarly to \cite{Fidkowski:2003nf} --see also \cite{Amado:2008hw}). 

As discussed above, in principle the free action can be supplemented by corrections arising from higher terms in the reduction of the SUGRA fluctuations as well as from higher curvature terms. Motivated by this, in this paper, concentrating on scalar operators, we consider a toy model which includes both a cubic bulk interaction between the bulk scalar fields as well as coupling to $W^2$. Assuming at least one neutral operator, the leading such coupling would be $\phi_k W^2$. This gives rise to a Witten diagram expansion as in fig.\eqref{2-pointGENERIC} below.\footnote{We could include as well the $|\phi|^2W^2$ term, in particular to account for charged operators: this is including the contribution of \cite{Rodriguez-Gomez:2021pfh}.} As we will see, the correction in fig.\eqref{2-pointGENERIC} has a few interesting consequences. First of all, the part of the diagram probing the black hole is really a straight radial geodesic which couples to $W^2$, and realizes the mechanism in \cite{Grinberg:2020fdj}. As a consequence, this correction will have exactly the same sensitivity to the time of travel to the singularity. Another consequence is that the whole diagram is proportional to a thermal VEV, which appears in this fashion in the 2-point thermal function. Indeed, if one looks to the leading term in fig.\eqref{2-pointGENERIC} --the vanilla 2-point function-- one can explicitly see \cite{Rodriguez-Gomez:2021pfh} how it contains the contributions from the operators $T^n$ --all possible contractions of $n$ energy-momentum tensors, with spins $0,2,\cdots,2n$--, but there is no trace of other thermal VEVs. This is a consequence of large 't Hooft coupling, and going beyond by including the correction in fig.\eqref{2-pointGENERIC} allows thermal VEV's to appear in the 2-point function. To be precise, the subleading term in fig.\eqref{2-pointGENERIC} captures the contribution of the operators $T^nO_k$ to the OPE ($n=0$ being the VEV itself). Finally, as we will discuss, the correction in fig.\eqref{2-pointGENERIC} is actually the leading term for off-diagonal correlators (an issue recently discussed in \cite{Engelsoy:2021fbk}). Recall that at $T=0$ conformal symmetry sets to zero 2-point functions for operators of different conformal dimensions. At $T\ne 0$ --or else, in $S^1\times \mathbb{R}^{d-1}$-- this is not  true anymore, and 2-point functions for operators with different dimensions may be non-vanishing. As discussed in \cite{Rodriguez-Gomez:2021pfh} and reviewed here, at large 't Hooft coupling where the gravity dual is captured by free fields in the black hole geometry, off-diagonal correlators vanish. On the other hand, the second diagram in fig.\eqref{2-pointGENERIC} is non-vanishing, and thus captures these off-diagonal correlators, which are then directly proportional to the thermal VEV's (and, therefore, consistently, vanish in the $T\rightarrow 0$ limit).

The main star of our paper is the second diagram in fig.\eqref{2-pointGENERIC}. As we will see, the VEV of the ``operator which corresponds to the straight line" can be factored out, leaving behind effectively a computation very similar to that of a 3-point function, which can be studied in the euclidean signature. It is however a peculiar 3 point function, as one of the geodesics, instead of running to the boundary, is anchored at the tip of the cigar (where it would meet the rest of the geodesic exploring the interior of the BH due to the $W^2$ term). This pulls the cubic coupling (the point in red in fig.\eqref{2-pointGENERIC}) down such that, before the separation in the thermal circle of the operator insertions reaches $\frac{\beta}{2}$, it touches the tip of the cigar. This marks a ``phase transition" where the correlator is dominated by a configuration of two radially straight geodesics meeting at the tip of the cigar (and then joining with the part which explores the interior due to the $W^2$ coupling). Even though this phase transition looks sharp in the geodesic approximation, we conjecture that finite $\Delta$ effects would smooth it. It would be very interesting to study this aspect in detail. 

As mentioned, we will work in euclidean signature. For the leading term, the euclidean version of the correlation function is in terms of real geodesics which live in the cigar geometry.\footnote{In Lorentzian signature things are more complicated: the geodesics can explore inside the horizon and probe both boundaries of the eternal black hole \cite{Hubeny:2012ry}, encoding in a subtle way information about the singularity \cite{Fidkowski:2003nf,Festuccia:2005pi}. Moreover, they can have ``bulk cone" singularities \cite{Hubeny:2006yu} --see also \cite{Amado:2008hw,Erdmenger:2011aa,Dodelson:2020lal}.} As shown in \cite{Rodriguez-Gomez:2021pfh,Rodriguez-Gomez:2021mkk}, these precisely reproduce the expectations from CFT not only showing the expected structure in terms of Gegenbauer polynomials but also recovering the correct numerical values of the central charges. In turn, for the subleading term in fig.\eqref{2-pointGENERIC}, the $W^2$ term forces the geodesic to ``explore inside the horizon" \cite{Grinberg:2020fdj}. This happens in an interesting way, since, as discussed above, the computation breaks into two pieces: the computation of the ``3-point function", which can be done completely in the euclidean cigar geometry; times the computation of a VEV, which just as in \cite{Grinberg:2020fdj}, is necessarily dominated by the singularity. Note that, since off-diagonal correlators are captured by second diagram in fig.\eqref{2-pointGENERIC}, they ``explore inside the horizon". Thus, it would be very interesting to further study off-diagonal correlators, specially in connection to \cite{Engelsoy:2021fbk}.

This paper is organized as follows: in section \ref{generalities} we set up the holographic computation of 2-point functions in our toy model, whose leading terms are captured by the Witten diagram expansion in fig.\eqref{2-pointGENERIC}. In section \ref{leading} we review the results for the first diagram in fig.\eqref{2-pointGENERIC} following \cite{Rodriguez-Gomez:2021pfh,Rodriguez-Gomez:2021mkk}, including a detailed discussion of the vanishing of that diagram for off-diagonal correlators. In section \ref{subleading} we discuss the computation of the subleading term in fig.\eqref{2-pointGENERIC}. Eq.\eqref{SL3} reflects the factorization of the VEV --which is just as in  \cite{Grinberg:2020fdj}-- and the ``3-point function" contribution. We then study explicitly the ``3-point" part (some relevant aspects of geodesics are compiled in appendix \ref{appendix:geodesics}). For diagonal correlators in the ``hierarchy" when the ``vertical field" is dual to an operator $O_k$ of a much smaller dimension, we can find a closed formula for the 2-point function where the contributing operators can be matched to $T^nO_k$. We then study the $d=2$ case in detail. Even though that case is a bit subtle\footnote{In this case $W^2$ vanishes. Conversely, since $S^1\times R$ is conformal to $R^2$, the exact 2-point function can be computed \cite{Cardy:1984rp}, showing that thermal VEVs, other than those for $T^n$, are not generated in $S^1\times R$. However, for the theory at finite volume VEV's can be generated through a bulk cubic coupling with a geodesic going around the black hole \cite{Kraus:2016nwo}.}, these subtleties go in the VEV pre-factor. In turn, we will use the $d=2$ ``3-point" piece as a toy model for other dimensions, as in $d=2$ the equations greatly simplify. This will allow us to discover the announced ``phase transition". We will then consider the $d=4$ case. In the OPE regime we will explicitly see how we recover the expected Gegenbauer polynomial structure, while for large separation of the insertions in the thermal circle, we will see that the same transition as in $d=2$ is at play.

\section{Holographic thermal correlators for neutral operators, including corrections}\label{generalities}

Let us consider the sector of scalar operators $\{O_i\}$ in a holographic CFT in $d$ dimensions. For simplicity, let us suppose that they are all are neutral under all would-be global symmetries which the CFT may have. In a top-down construction, the bulk action would follow from reducing the appropriate String Theory on the relevant internal space, leaving behind an effective theory for the fluctuations in a $d+1$ dimensional asymptotically $AdS_{d+1}$ space containing a black hole. In the strict limit of large central charge $c_{\mathcal{T}}$ and large coupling $\lambda$ (large $N$ and large 't Hooft coupling in the paradigmatic $\mathcal{N}=4$ SYM example), the gravity dual is in terms of supergravity fluctuations in the geometry of the black brane in $AdS_{d+1}$. One may imagine to systematically include $c_{\mathcal{T}}^{-1}$ and $\lambda^{-1}$ effects, which would correspond to  higher order terms in the effective action for the fluctuations and  higher curvature corrections to the background. Thus, on general grounds we are led to consider the bulk action

\begin{equation}
\label{bulkaction}
S=\int_{\rm bulk} \sqrt{g}\,\Big(\frac{1}{2}(\partial\phi_i)^2+\frac{1}{2}m_i^2\phi_i^2+\lambda_{ijk}\phi_i\phi_j\phi_k+\alpha_i\,\phi_i\,W^2\Big)\,,
\end{equation}
where the metric is given by the appropriate correction to the usual black brane in $AdS_{d+1}$ (for instance, such metric was explicitly constructed in \cite{Gubser:1998nz} for the paradigmatic $\mathcal{N}=4$ SYM example), and $W$ is its Weyl tensor.

\begin{figure}[h!]
\centering
\includegraphics[scale=.2]{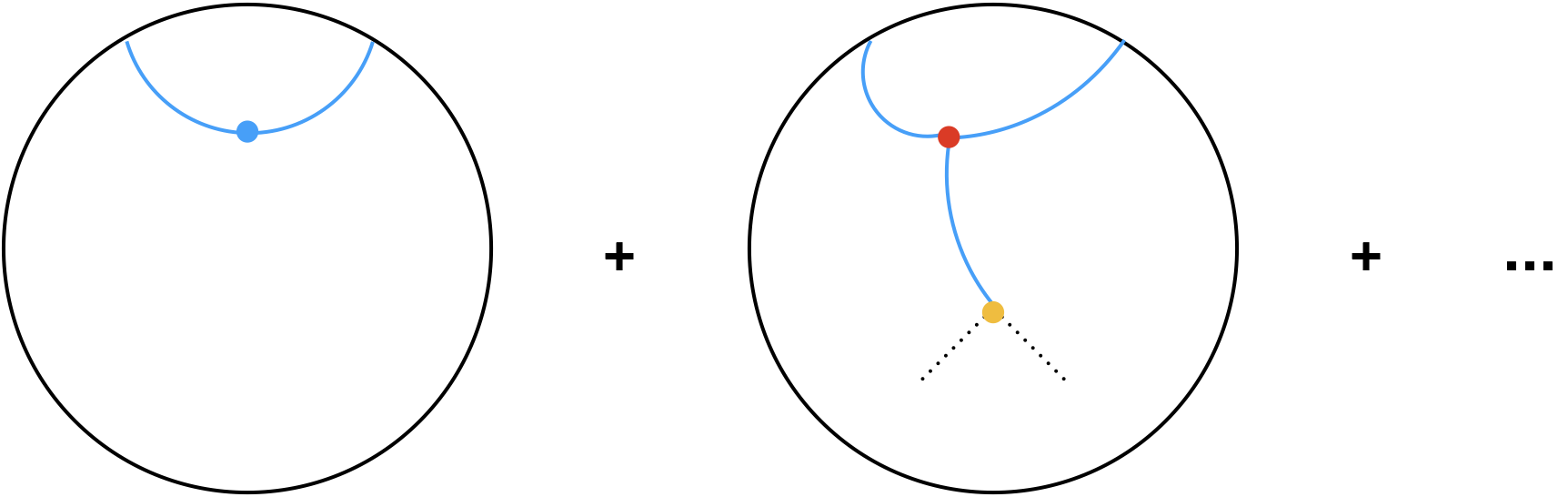}
\caption{Expansion in Witten diagrams of the 2-point function. The blue vertex represents the ``trivial vertex" breaking in two the geodesic for the 2-point function. The red vertex corresponds to the cubic interaction $\lambda_{ijk}$. The yellow vertex corresponds to the higher curvature coupling to $W^2$.}
\label{2-pointGENERIC}
\end{figure}

Let us first ignore the cubic coupling as well as the coupling to $W^2$. The resulting action allows to holographically compute the 2-point function for the operators $O_i$. As usual, one would solve the equation of motion with the appropriate boundary conditions, from which the desired 2-point function in the boundary theory can be read-off. A particularly simple sector is that of operators of large dimension. For $\Delta_i\gg 1$ (but still, smaller than $c_{\mathcal{T}}$), $m_i\sim \Delta_i$, and thus the bulk equation of motion can be solved in the semiclassical WKB approximation. One can easily show that the WKB solution is essentially the exponential of the action for a particle of mass $m_i$  travelling through the bulk from the insertion point of one operator to the insertion point of the other operator, that is, the exponential of the geodesic length between the insertion points of the operators. Moreover, it turns out that one can regard this geodesic as the junction at some bulk point $x_I=(\tau_I,\vec{x}_I,z_I)$ of two geodesic arcs upon integrating over the bulk the junction point. This allows to regard the 2-point function as an interacting Witten diagram. Then, including the cubic interaction and the coupling to $W^2$ results in more vertices leading to more Witten diagrams contributing to the desired 2-point function. To leading order we would find as in fig.\eqref{2-pointGENERIC}.

Let us denote by $G^{(i)}_{\partial \rm{b}}(x,u)$ the bulk-to-boundary propagator from a bulk point $x$ to a boundary point $u$ for the bulk field $i$ --dual to the operator $i$, and hence with mass $m_i$. In addition, $G^{(i)}_{\rm{b} \rm{b}}(x,y)$ will be the bulk-to-bulk propagator for the bulk field $i$ from $x$ to $y$. Then, the 2-point function resulting from fig.\eqref{2-pointGENERIC} would be

\begin{eqnarray}
\label{2-pointGENERIC-formula}
\langle O_i(x_1)\,O_j(x_2)\rangle&=&\int_{\rm bulk}du\,G^{(i)}_{\partial \rm{b}}(x_1,u)\,G^{(j)}_{\partial \rm{b}}(x_2,u)\\ \nonumber && +\lambda_{ijk}\,\alpha_k\,\int_{\rm bulk}du\,\int_{\rm bulk} dv\, G^{(i)}_{\partial \rm{b}}(x_1,u)\,G^{(j)}_{\partial \rm{b}}(x_2,u)\,G^{(j)}_{\rm{b} \rm{b}}(u,v)\,W^2(v)+\cdots\,.
\end{eqnarray}

In the large dimension limit, as described above, both the bulk-to-boundary and the bulk-to-bulk propagators are proportional to the exponential of the geodesic length between the arguments. Thus, in the following we will suppress the subscripts and just write $G^{(i)}(x,y)\sim e^{-S_i}$, being $S_i$ the action for a mass $m_i$ particle whose trajectory joints $x$ and $y$. As we are interested in including corrections, $G$ is to be computed with the appropriately $\alpha'$ corrected black brane background. Note however that, since $\alpha_i$ in \eqref{bulkaction} is already proportional to $\alpha'$, to leading order the second line in \eqref{2-pointGENERIC-formula} can be evaluated in the zeroth-order black-brane background, which reads (in lorentizan signature)

\begin{equation}
\label{blackbrane}
ds^2=\frac{R^2}{z^2}\Big(-f(z)dt^2+\frac{dz^2}{f(z)}+d\vec{x}_{d-1}^2\Big)\,,\qquad f(z)=1-\frac{z^d}{z_0^d}\,,\qquad z_0=\frac{d}{4\pi}\beta\,.
\end{equation}
We will sometimes use units in which $z_0=1$. Re-storing factors of temperature is straightforward on dimensional grounds.

\section{The leading term}\label{leading}

Let us consider the leading term in \eqref{2-pointGENERIC-formula}. Its contribution reads

\begin{equation}
\label{L}
\langle O_i(x_1)\,O_j(x_2)\rangle_L=\int_{\rm bulk}du\,G^{(i)}_{\partial \rm{b}}(x_1,u)\,G^{(j)}_{\partial \rm{b}}(x_2,u)\,.
\end{equation}
As described above, we would like to evaluate, in the large dimension approximation, the contribution \eqref{L} including the first $\alpha'$ correction. Nevertheless, let us first consider the ``tree level" background, given by the black brane geometry in \eqref{blackbrane}. Then, the necessary ingredients for the computation of this diagram have been very recently developed in \cite{Rodriguez-Gomez:2021pfh,Rodriguez-Gomez:2021mkk}. One finds that $G^{(i)}(x,y)\sim e^{-\Delta_i\,S_{\rm arc}(x,y)}$ for some function $S_{\rm arc}(x,y)$ --actually the length of a geodesic arc between $x$ and $y$-- which is discussed in those references, and briefly reviewed in the appendix. Thus \eqref{L} boils down to 

\begin{equation}
\langle O_i(x_1)\,O_j(x_2)\rangle_L=\int_{\rm bulk}du\,e^{-S}\,\qquad S=\Delta_i\,S_{\rm arc}(x_1,u)+\Delta_j\,S_{\rm arc}(x_2,u)\,.
\end{equation}
Since $\Delta_i\gg 1$, we can perform the integral above in the saddle point approximation, which boils down to evaluating the integrand in the solution to the saddle point equations $\frac{dS}{du}=0$ (here $u=(\tau,\vec{x},z)$). We can read off the contribution of each geodesic segments to these equations from \cite{Rodriguez-Gomez:2021mkk}. These equations boil down to

\begin{eqnarray}
\label{Lsaddle}
&&\Delta_i\,\sqrt{f(z)-z^2\mu_i^2+f(z)\,z^2\,\nu_i^2}\pm \Delta_j\,\sqrt{f(z)-z^2\mu_j^2+f(z)\,z^2\,\nu_j^2}=0\,;\\ \nonumber && \Delta_i\mu_i+\Delta_j\mu_j=0\,,\qquad \Delta_i\nu_i+\Delta_j\nu_j=0\,;
\end{eqnarray}
where $\mu_{i,j},\nu_{i,j}$ are the momenta labelling the geodesics (see \cite{Rodriguez-Gomez:2021pfh,Rodriguez-Gomez:2021mkk} and the appendix for details), and $\pm$ refers to the fact that one may wonder whether one should use ingoing/ingoing or ingoing/returning geodesics (see \cite{Rodriguez-Gomez:2021mkk} and the appendix for details). One can check that if $\Delta_i\ne \Delta_j$, the only solution to these equations is $z_I=z_0$ and $\mu_i=\mu_j=0$, $\nu_i=\nu_j=0$. This corresponds to ``straight" geodesics running radially at fixed $\tau$ so that the resulting configuration of joining two of them would have a ``wedge" and hence would result in a non-smooth curve --unless the insertions are opposite located in the circle, so that the two ``straight geodesics" are aligned. As a result, there is no admissible saddle configuration if $\Delta_i\ne \Delta_j$, which shows that the leading contribution to the off-diagonal correlator vanishes. Note that, leaving aside the ``wedge", the $z_I=z_0$ configuration would lead to rather bizarre features for the correlator, such as a nonsensical $T\rightarrow 0$ limit, or the fact that, since $\mu_i=\mu_j=\nu_i=\nu_j=0$, the correlator would not depend at all on spacetime.

Let us now turn on the first $\alpha'$ correction. At low temperatures (more precisely, at small $T|x|$), and in $d=4$, this was considered,using the background in \cite{Gubser:1998nz}, in \cite{Rodriguez-Gomez:2021mkk}; where it was argued that the leading correction only changes the coefficients by effectively shifting $T^4\rightarrow T^4\,(1+\frac{15}{8}\zeta(3)\,(2\lambda)^{-\frac{3}{2}})$. The reason for this is that the first $\alpha'$ correction to the geometry \eqref{blackbrane}, for small $T$, results in a background of the same form of \eqref{blackbrane} itself only that $z_0$ is related in a different way to the temperature. As a consequence, the computation proceeds just as in the $\alpha'$ un-corrected case. In particular, the argument above should go through, and so $\langle O_i(x)\,O_j(0)\rangle_L\sim \delta_{ij}$ even including the first $\alpha'$ correction. One may wonder whether including higher orders this conclusion could change, and an off-diagonal correlator can be generated. To begin with, note that any case, the coupling to $W^2$ would have less $\alpha'$ suppression. Hence, the subleading contribution to be analysed in the next section would still be the leading one. However, that higher $\alpha'$ corrections to the background generate contributions to off-diagonal correlators looks unlikely, as the higher corrections in \cite{Gubser:1998nz} arise as an expansion of the background in powers of $z^{-4}$. This has the right structure so as to correct the contributions of $T^n$ to the correlator rather than generating new contributions.

Thus, all in all, borrowing the results from \cite{Rodriguez-Gomez:2021pfh,Rodriguez-Gomez:2021mkk}, the contribution from the leading term to the correlator is

\begin{equation}
\label{leadingterm}
\langle O_i(x)\,O_j(0)\rangle_L=\frac{\delta_{ij}}{|x|^{2\Delta_i}}\,e^{A_d\,T^d\,|x|^d\,C_2^{(\frac{d-2}{2})}(\eta)+\mathcal{O}((T^d|x|^d)^2)}\,,\qquad A_d=\frac{2^{d-2}\,\pi^{\frac{2d+1}{2}}\,\Gamma\big(\frac{d-2}{2}\big)}{d^d\,\Gamma\big(\frac{d+3}{2}\big)}\,(1+\delta A_d)\,;
\end{equation}
where we have introduced $|x|=\sqrt{\tau^2+\vec{x}^2}$ and $\eta=\frac{\tau}{|x|}$, and where $C_2^{(\frac{d-2}{2})}(\eta)$ is the corresponding Gegenbauer polynomial. The correction term $\delta A_d$ is of the order $\alpha'^3$, and for $d=4$ reads $\delta A_4=\frac{15}{8}\zeta(3)(2\lambda)^{-\frac{3}{2}}$.

\section{The subleading term}\label{subleading}

Let us now go to the subleading contribution in \eqref{2-pointGENERIC-formula}, which is given by

\begin{equation}
\label{SL}
\langle O_i(x_1)\,O_j(x_2)\rangle_{SL}=\lambda_{ijk}\,\alpha_k\,\int_{\rm bulk}du\,\int_{\rm bulk} dv\, G^{(i)}_{\partial \rm{b}}(x_1,u)\,G^{(j)}_{\partial \rm{b}}(x_2,u)\,G^{(j)}_{\rm{b} \rm{b}}(u,v)\,W^2(v)\,.
\end{equation}
Recall that, since $\alpha_k$ is already of the order of $\alpha'$, to leading order we should compute \eqref{SL} in the black brane background \eqref{blackbrane}. Moreover, we will also assume  $\Delta_k\gg 1$. Then, once again, in the large dimension approximation we can approximate both bulk and boundary propagators by $G(x,y)\sim e^{-\Delta\,S_{\rm arc}(x,y)}$. Moreover, it turns out that when evaluated in the black brane background 

\begin{equation}
W^2=\frac{d\,(d-2)\,(d-1)^2}{R^2}\,\frac{z^d}{z_0^d}\,.
\end{equation}
Thus, we may write 

\begin{eqnarray}
\label{SL2}
\langle O_i(x_1)\,O_j(x_2)\rangle_{SL}&=&\frac{d\,(d-2)\,(d-1)^2}{R^2}\,\lambda_{ijk}\,\alpha_k\\ \nonumber &&\times \int_{\rm bulk}du\,e^{-\Delta_i\,S_{\rm arc}(x_1,u)-\Delta_j\,S_{\rm arc}(x_2,u)}\,\int_{\rm bulk} dv\,e^{-\Delta_k\,S_{\rm arc}(u,v)-d\,\log\frac{v_z}{z_0}}\,,
\end{eqnarray}
where we use the notation that $v_z$ parametrizes the radial coordinate of the bulk point $v$.

In order to simplify the otherwise undoable $v$ integral, let us approximate it by a saddle point approximation.\footnote{One way to justify this would be to consider the limit $d\sim \Delta_k$.} Denoting by $v_t$ and $\vec{v}_x$ the time and spatial components of $v$, the saddle point equations would read

\begin{equation}
\label{saddleVEV}
\frac{dS_{\rm arc}(u,v)}{dv_t}=0\,,\qquad \frac{dS_{\rm arc}(u,v)}{d\vec{v}_x}=0\,,\qquad \Delta_k\,\frac{dS_{\rm arc}(u,v)}{dv_z}+\frac{d}{v_z}=0\,.
\end{equation}

As shown in \cite{Rodriguez-Gomez:2021mkk} (and reviewed in the appendix), $\frac{dS_{\rm arc}(u,v)}{dv_t}=\mu$ and $\frac{dS_{\rm arc}(u,v)}{d\vec{v}_x}=\nu$. Since the first two equations in \eqref{saddleVEV} have no compensating term, they set $\mu=\nu=0$ for the geodesic connecting the $u$ and $v$ points. This implies that such geodesic is vertical --that is, along the radial direction--, and thus its contribution is very simple

\begin{equation}
S_{\rm arc}(u,v)=\int_{u_z}^{v_z}\,dz \frac{1}{z\,\sqrt{f(z)}}\,.
\end{equation}
We may write this as follows

\begin{equation}
S_{\rm arc}(u,v)=\int_{0}^{v_z}\,dz \frac{1}{z\,\sqrt{f(z)}}-\int_{0}^{u_z}\,dz \frac{1}{z\,\sqrt{f(z)}}\,.
\end{equation}
The first integral is nothing but the geodesic ``vertical" length $\ell$ between the boundary and a point at depth $v_z$. Thus, we may write \eqref{SL2} as

\begin{equation}
\label{SL3}
\langle O_i(x_1)\,O_j(x_2)\rangle_{SL}=\lambda_{ijk}\,\int_{\rm bulk}du\,e^{-\Delta_i\,S_{\rm arc}(x_1,u)-\Delta_j\,S_{\rm arc}(x_2,u)+\Delta_k\,S_{\rm vert}(0,u)}\,\Big[ \alpha_k\,\int_{\rm bulk} dv\,e^{-\Delta_k\,\ell}\,W^2\Big]\,,
\end{equation}
where

\begin{equation}
S_{\rm vert}(0,u)=\int_{0}^{u_z}\,dz \frac{1}{z\,\sqrt{f(z)}}\,.
\end{equation}

The quantity in brackets in \eqref{SL3} does not depend on $u$, and thus can be extracted from the integral. In fact, that piece is nothing but the thermal VEV of the $O_k$ operator \cite{Grinberg:2020fdj}. Thus, we can write

\begin{equation}
\label{SL3}
\langle O_i(x_1)\,O_j(x_2)\rangle_{SL}=\langle O_k\rangle\,\lambda_{ijk}\,\int_{\rm bulk}du\,e^{-\Delta_i\,S_{\rm arc}(x_1,u)-\Delta_j\,S_{\rm arc}(x_2,u)+\Delta_k\,S_{\rm vert}(0,u)}\,.
\end{equation}

The integral in \eqref{SL3} needs not to vanish for $i\ne j$. Thus, since the leading term in \eqref{leadingterm} is proportional to $\delta_{ij}$, \eqref{SL3} actually provides the leading contribution to non-diagonal 2-point functions. As expected, it is only non-zero due to the fact that in the thermal background $\langle O_k\rangle\ne 0$.  

Interestingly, \eqref{SL3} factors out the VEV --which corresponds to a vertical geodesic all the way to the singularity due to the $W^2$ term as in \cite{Grinberg:2020fdj}--  times a factor which is very reminiscent of a 3-point function (see \cite{Rodriguez-Gomez:2021mkk}), only that the vertical segment, instead of running to the boundary, is anchored at the tip of the cigar at $z=z_0$ (morally speaking, from there on it joins the vertical piece probing the interior).

In order to further proceed, let us explicitly evaluate $S_{\rm vert}$. Introducing a boundary regulator $\epsilon$, one finds ($u=(t,\vec{x},z))$

\begin{equation}
S_{\rm vert}=\frac{1}{d}\,\log\Big( \frac{1-\sqrt{f(z)}}{1+\sqrt{f(z)}}\Big)-\log\frac{\epsilon}{z_0\,4^{\frac{1}{d}}}\,.
\end{equation}
Note that

\begin{equation}
\frac{dS_{\rm ver}}{dz}=\frac{1}{z\,\sqrt{f(z)}}\,.
\end{equation}
Thus, the saddle point equations are

\begin{align}
\label{saddleeq}
&\Delta_i\,\mu_i+\Delta_j\,\mu_j=0\,,\qquad \Delta_i\,\nu_i+\Delta_j\,\nu_j=0\,, \\ \nonumber & \\& \Delta_i\,\sqrt{f(z)-z^2\,\mu_i^2-f(z)\,z^2\,\nu_i^2}+ \Delta_j\,\sqrt{f(z)-z^2\,\mu_j^2-f(z)\,z^2\,\nu_j^2}-\Delta_k\,\sqrt{f(z)}=0\,. 
\end{align}

Let us consider the case of diagonal correlators $\Delta_i=\Delta_j$ at coincident spatial points, corresponding to $\nu_{i,j}=0$. Then the saddle equations become

\begin{equation}
\label{saddleeqtime}
\big(1-\frac{\Delta_k^2}{4\Delta_i^2}\big)\,(1-\frac{z^d}{z_0^d})=z^2\,\mu_i^2\,; 
\end{equation}
Then, for $\mu_i\in[0,\infty)$ one has $z\in[z_0,0)$, that is, the bulk integral in \eqref{SL3} lives in the euclidean part of the geometry corresponding to ``outside the horizon", while the one-point VEV does probe the interior of the black hole as described in \cite{Grinberg:2020fdj}. Note that this is for $\Delta_k<2\Delta_i$. For the extremal case $\Delta_k=2\Delta_i$ the saddle goes to the boundary and the correction splits up into the product of two 2-point functions, just as in \cite{Rodriguez-Gomez:2021mkk}.

\subsection{Diagonal correlators in the hierarchy $\Delta_i=\Delta_j=\Delta\gg \Delta_k$}

It is interesting to consider diagonal correlators $\Delta_i=\Delta_j=\Delta$ in the hierarchy of dimensions $\Delta\gg \Delta_k\gg 1$. In that case the $\Delta_k$ term in \eqref{SL3} does not contribute to the saddle point equation, and we can write, for all practical purposes

\begin{equation}
\langle O_i(x_1)\,O_i(x_2)\rangle_{SL}=\langle O_k\rangle\,\lambda_{iik}\,\langle O_i(x_1)\,O_i(x_2)\rangle_{L}\,\Big(e^{\Delta_k\,S_{\rm vert}(0,u)}\Big|_{u_{2}}\Big)\,;
\end{equation}
where $u_2$ is the solution to the saddle equation for the 2 point correlator \eqref{Lsaddle}. Concentrating on equal-time correlators by setting $\nu=0$, we can borrow the result from eq.(3.6) in \cite{Rodriguez-Gomez:2021mkk}, so that

\begin{eqnarray}
\label{Dk<<D}
\langle O_i(\tau)\,O_i(0)\rangle_{SL}&=&\langle O_k\rangle\,\lambda_{iik}\,\langle O_i(\tau)\,O_i(0)\rangle_{L}\,\tau^{\Delta_k}\, e^{\Delta_k\,c_d\,(T\,\tau)^d+\mathcal{O}\big((T\tau)^{2d}\big)}\,,\nonumber\\ c_d&=&\frac{1}{2d}\,\Big(\frac{2\pi}{d}\Big)^d\,\Big(1-d+\frac{(d-1)\,d\,\sqrt{\pi}\,\Gamma(1+\frac{d}{2})}{2\,\Gamma(\frac{3+d}{2})}\Big)\,.
\end{eqnarray}
Let us expand the exponential, and use that, to leading order, we can write $\langle O_i(\tau)\,O_i(0)\rangle_{L}=\tau^{-2\Delta}$. One finds

\begin{equation}
\langle O_i(\tau)\,O_i(0)\rangle_{SL}=\langle O_k\rangle\,\lambda_{iik}\,\tau^{\Delta_k-2\Delta}+\frac{\Delta_k\,c_d\,\langle O_k\rangle\,\lambda_{iik}}{\beta^d}\,\tau^{\Delta_k+d-2\Delta}+\cdots\,.
\end{equation}
It is natural to regard this as the $|\vec{x}|\rightarrow 0$ limit of (this will be justified explicitly in $d=4$ in section \eqref{spatial} below)

\begin{equation}
\langle O_i(\tau)\,O_i(0)\rangle_{SL}=\langle O_k\rangle\,\lambda_{iik}\,C_0^{(\frac{d}{2}-1)}(\eta)\,|x|^{\Delta_k-2\Delta}+\frac{\Delta_k\,c_d\,\langle O_k\rangle\,\lambda_{iik}}{\beta^d}\,C_2^{(\frac{d}{2}-1)}(\eta)\,|x|^{\Delta_k+d-2\Delta}+\cdots\,;
\end{equation}
where $C_J^{(\nu)}(\eta)$ are the Gegenbauer polynomials of $\eta=\frac{\tau}{|x|}$, $|x|=\sqrt{\tau^2+|\vec{x}|^2}$ and we assume $d>2$. We recognize here the structure in \cite{Iliesiu:2018fao}, corresponding each term to the contribution of the scalar operator $\langle O_k\rangle$ of dimension $\Delta_k$ and of the spin 2 and dimension $d+\Delta_k$ operator $T\,O_k$ --$T$ being the energy-momentum tensor. More generically, expanding the exponential and re-writting it in terms of Gegenbauer polynomials suggests that the subleading contribution captures the contributions of $T^nO_k$, where $T^n$ stands for all possible contractions of $n$ energy-momentum tensors with spin $0,2,\cdots,2n$. Note that the coefficient of the $n=0$ term, which the contribution of the $O_k$ operator itself, matches the expectation that the coefficient in the 2-point function must be precisely $\langle O_k\rangle \lambda_{iik}$. As for the second, it constitutes a prediction for the product of the VEV of the $TO_k$ operator --denoted $b_{(TO_k)}$-- times the OPE coefficient $O$-$O$-$(TO_k)$ --denoted by $\lambda_{OO(TO_k)}$-- and all normalized to its 2-point function --denoted $c_{(TO_k)}$--:

\begin{equation}
    \frac{\lambda_{OO(TO_k)}\,b_{(TO_k)}}{c_{(TO_k)}}=\Delta_k\,b_{O_k}\,\lambda_{iik}\,\frac{(d-2)}{4}\,\Big(\frac{2\pi}{d}\Big)^d\,\Big(1-d+\frac{(d-1)\,d\,\sqrt{\pi}\,\Gamma(1+\frac{d}{2})}{2\,\Gamma(\frac{3+d}{2})}\Big)\,,
\end{equation}
where $b_{O_k}=\langle O_k\rangle\,\beta^{\Delta_k}$.

\subsection{Time-dependent correlators in $d=2$}

Let us restrict to the case of $d=2$. Note that in $d=2$, $W^2$ really vanishes. In particular, this implies that off-diagonal correlators vanish in $d=2$ \cite{Iliesiu:2018fao}. However, for the finite-area BTZ case it is possible to generate a VEV through the 3-point coupling to a geodesic ``wrapping the horizon" \cite{Kraus:2016nwo}. This will go into the details of $\langle O\rangle$, but as far as the rest of the computation is concerned, we can simply extend our previous formulas to $d=2$, where they simplify so as to allow for an explicit treatment serving as a toy model for higher dimensions. Moreover, we will restrict to equal $x$ correlators. Then, the saddle equations set $\nu_{1,2}=0$ and

\begin{equation}
\mu_j=-\frac{\Delta_i\,\mu_i}{\Delta_j}\,.
\end{equation}
Then

\begin{equation}
\label{z2d}
z_I=\frac{1}{\sqrt{1+\frac{4\Delta_i^2\Delta_k^2}{[\Delta_k^2-(\Delta_i-\Delta_j)^2]\,[(\Delta_i+\Delta_j)^2-\Delta_k^2]}\mu_i^2}}\,.
\end{equation}
Note that $z\in \mathbb{R}$ provided

\begin{equation}
\label{range2d}
|\Delta_i-\Delta_j|\leq \Delta_k\leq |\Delta_i+\Delta_j|\,.
\end{equation}

Moreover, evaluating the equations of motion at \eqref{z2d} gives the intersection point. Restricting to the region of small $\tau$, one can solve for $\mu_i$ as

\begin{equation}
    \mu_i^{-1}=\frac{\Delta_i}{\Delta_i+\Delta_j-\Delta_k}\,\tau+\frac{\Delta_i\,(\Delta_k^2+2\Delta_k\,(\Delta_i+\Delta_j)-3(\Delta_i-\Delta_j)^2)}{24\,(\Delta_i+\Delta_j-\Delta_k)^2\,\Delta_k}\,\tau^3+\cdots\,.
\end{equation}
Plugging this into \eqref{correlator2dmu} one finds (we neglect numerical factors, not relevant for our purposes now)

\begin{equation}
    \langle O_i(\tau)\,O_j(0)\rangle_{SL}\sim \langle O_k\rangle \lambda_{ijk}\,\frac{e^{4\pi^2\,\frac{\Delta_k^2+2\Delta_k (\Delta_i+\Delta_j)-3(\Delta_i-\Delta_j)^2}{48\Delta_k}\,(T\tau)^2+\cdots}}{\tau^{\Delta_i+\Delta_j-\Delta_k}}\,.    
\end{equation}
To cross-check this equation, setting $\Delta_i=\Delta_j=\Delta$ gives

\begin{equation}
    \langle O(\tau)\,O(0)\rangle_{SL}\sim \langle O_k\rangle \lambda_{iik}\,\frac{e^{\Delta\,\frac{\pi^2}{3}\,(T\tau)^2}}{\tau^{2\Delta}}\,\tau^{\Delta_k}\,e^{\Delta_k\frac{\pi^2}{12}\,(T\tau)^2+\cdots}\,.    
\end{equation}
We recognize the leading 2-point function for $O_i$ (\textit{c.f.} eq.(3.34) in \cite{Rodriguez-Gomez:2021pfh}), while the second part reproduces the $d=2$ case of \eqref{Dk<<D}.

\subsubsection{The correlator in the whole range of $\tau$}

Let us now investigate the correlator for arbitrary $\tau\in [0,\beta]$. To simplify the analysis, let us consider the case $\Delta_i=\Delta_j=\Delta_k\equiv \Delta$. Due to the equal $\Delta$ restriction (specifically, due to $\Delta_i=\Delta_j$) the resulting geodesic arrangement is symmetric, following from the fact that in this case the saddle point equations demand $\mu\equiv\mu_i=-\mu_j$. Evaluating the action, one finds

\begin{equation}
\label{correlator2dmu}
    \langle O_i(\tau)\,O_i(0)\rangle_{SL}=\langle O_i\rangle \lambda_{iii} \Big(-\frac{|\mu|^3}{6\sqrt{3}}+\frac{3+\mu^2}{12\sqrt{3}}\,\sqrt{3+4\mu^2}\Big)^{\Delta}\,.
\end{equation}

Moreover, evaluating the equations of motion at \eqref{z2d} gives the intersection point. One finds

\begin{equation}
    i\tau_I=i\tau_i+\frac{1}{2}\,\log\Big( \frac{(-i+\mu)\,(i\mu-\frac{|\mu|}{\sqrt{3+4\mu^2}})}{(i+\mu)(i\mu+\frac{|\mu|}{\sqrt{3+4\mu^2}}}\Big)\,;\qquad   i\tau_I=i\tau_j-\frac{1}{2}\,\log\Big( \frac{(-i+\mu)\,(i\mu-\frac{|\mu|}{\sqrt{3+4\mu^2}})}{(i+\mu)(i\mu+\frac{|\mu|}{\sqrt{3+4\mu^2}}}\Big)\,.
\end{equation}
Equating these, that is, demanding that the two geodesic arms corresponding to $O_{i,j}$ meet at \eqref{z2d}, allows to fix $\mu$ as a function of $\tau=\tau_1-\tau_2$ one finds 

\begin{equation}
\label{eq2d}
    i\tau+\log\Big( \frac{(-i+\mu)\,(i\mu-\frac{|\mu|}{\sqrt{3+4\mu^2}})}{(i+\mu)\,(i\mu+\frac{|\mu|}{\sqrt{3+4\mu^2}}}\Big)=0\,.
\end{equation}
The argument of the logarithm is really a modulus one complex number. Hence, $\theta$ is defined modulo $2\pi$ and so is $\tau$. We thus find the invariance $\tau\rightarrow \tau+2\pi$. Recovering factors of temperature, this is the familiar KMS condition $\tau\rightarrow \tau+\beta$.

The solution to \eqref{eq2d} is

\begin{equation}
    \mu=i\frac{1+e^{i\tau}-\sqrt{e^{i\tau}}}{1-e^{i\tau}}\,.
\end{equation}
This function is

\begin{equation}
\label{brutemu}
    \mu=\begin{cases} -\frac{1}{\sin\frac{\tau}{2}}\,\Big(\cos\frac{\tau}{2}-\frac{1}{2}\Big)\qquad \tau\in [0,\pi]\,,\\ \\
   -\frac{1}{\sin\frac{\tau}{2}}\,\Big(\cos\frac{\tau}{2}+\frac{1}{2}\Big)\quad \tau\in (\pi,2\pi]\,.
    \end{cases}
\end{equation}
However, one can check --for instance numerically-- that, when plugged back in the original equation this is a solution only in the interval $\tau\in[0,\frac{2\pi}{3}]\cup [\frac{4\pi}{3},2\pi]$ (of course, the dimensions are hidden in a factor $z_0^{-1}$ not written as we are taking $z_0=1$).  Thus, this configuration does not exist in the central interval $\tau\in[\frac{2\pi}{3},\frac{4\pi}{3}]$. In order to figure out this central interval, note first that at the borders $\tau=\frac{2\pi}{3},\frac{4\pi}{3}$, eq.\eqref{brutemu} gives $\mu=0$, which corresponds to straight geodesics going all the way to $z=1$. Motivated by this, consider now a straight geodesic going, at fixed $\tau$, from the boundary all the way to $z=1$. One could imagine a configuration of two such straight geodesics meeting with the vertical segment at $z=1$ (let's call this configuration the ``straight" configuration as opposed to the standard ``U-shaped" one\footnote{Which is not really ``U-shaped" as we are considering the subleading correction to the 2-point function coming from a diagram with a cubic vertex.}). Note that, as discussed for the leading term, the straight configuration has a ``wedge" at the bottom unless $\tau=\frac{\beta}{2}$. However, in the case at hand, as we are evaluating a genuine Witten diagram in the geodesic approximation, the resulting full configuration needs not to be a smooth geodesic itself. Now, since at $\tau=\frac{2\pi}{3},\frac{4\pi}{3}$ the U-shaped configuration hits $z=1$ with $\mu=0$, at that point it actually becomes a straight configuration. Hence, it is natural to conjecture that from there on and in the whole middle interval, it is the straight configuration what describes the correlator in the central interval $\tau\in[\frac{2\pi}{3},\frac{4\pi}{3}]$. Thus

\begin{equation}
\label{mu}
    \mu=\begin{cases} -\frac{1}{\sin\frac{\tau}{2}}\,\Big(\cos\frac{\tau}{2}-\frac{1}{2}\Big)\qquad \tau\in [0,\frac{2\pi}{3}]\,,\\  \\ 0 \qquad \tau\in[\frac{2\pi}{3},\frac{4\pi}{3}]\,, \\ \\
   -\frac{1}{\sin\frac{\tau}{2}}\,\Big(\cos\frac{\tau}{2}+\frac{1}{2}\Big)\quad \tau\in [\frac{4\pi}{3},2\pi]\,.
    \end{cases}
\end{equation}
A cartoon of the resulting configuration is in fig.\eqref{cartoon2d} below.

\begin{figure}[h!]
    \centering
          \includegraphics[scale=.25]{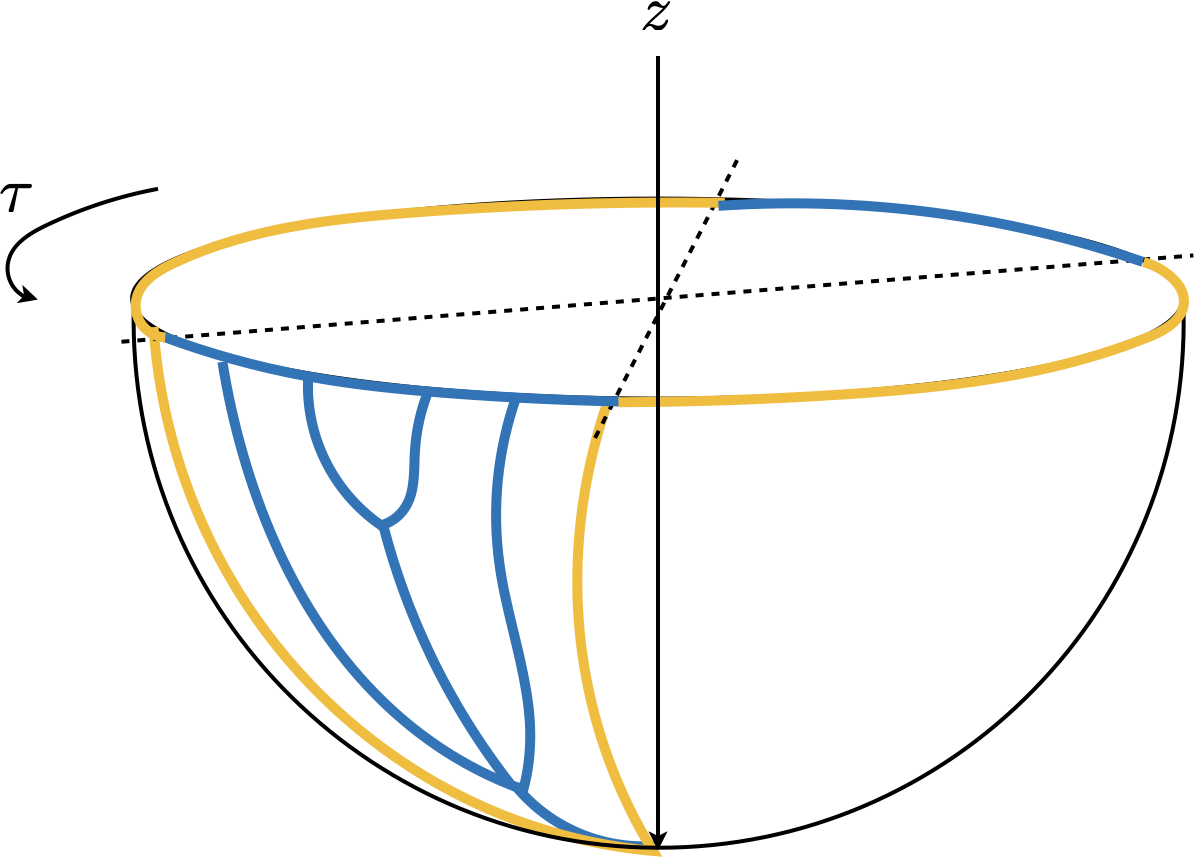}
          \caption{Cartoon of the dominating configurations: as the boundary time separation increases, the U penetrates more and more into the bulk pulled by the vertical segment (blue configurations in blue zone). Eventually the tip of the U hits $z=1$ and the only configuration is the straight one (yellow configurations in yellow zone).}
          \label{cartoon2d}
\end{figure}

Note that the $\mu$ above is continuous at $\tau=\frac{2\pi}{3},\,\frac{4\pi}{3}$, which implies that the resulting correlator is a continuous function. However, it is not infinitely differentiable at those points. To see this, note that the correlator in eq.\eqref{correlator2dmu} depends on $\tau$ implicitly through $\mu$. Hence, using the chain rule, one has that taking the first derivative with respect to $\tau$ will give something proportional to $\mu$. Then, since that at the borders $\tau=\frac{2\pi}{3},\frac{4\pi}{3}$ we have $\mu=0$, the first derivative vanishes. On the other hand, to the other side where the correlator is given by the straight configuration, it is obvious that all derivatives, in particular the first, vanish, giving a continuous first derivative. However, taking the second derivative includes a term proportional to $\sqrt{3+4\mu^2}$, which at the borders $\tau=\frac{2\pi}{3},\frac{4\pi}{3}$ where $\mu=0$, does not vanish, making the second derivative not continuous. Hence, since the correlator is not differentiable infinitely many times at the borders $\tau=\frac{2\pi}{3},\frac{4\pi}{3}$, in a sense it exhibits a sharp ``phase transition" at those points. However, that this phase transition occurs sharply seems an artefact of the geodesic approximation: solving the full problem --\textit{i.e.} solving the KG equation in the black hole background to compute exactly the Witten diagram-- would smooth out the transition, making it of class $\mathcal{C}^{\infty}$. The correlator is show in fig.\eqref{correlator2d} below.

\begin{figure}[h!]
    \centering
      \includegraphics[scale=.5]{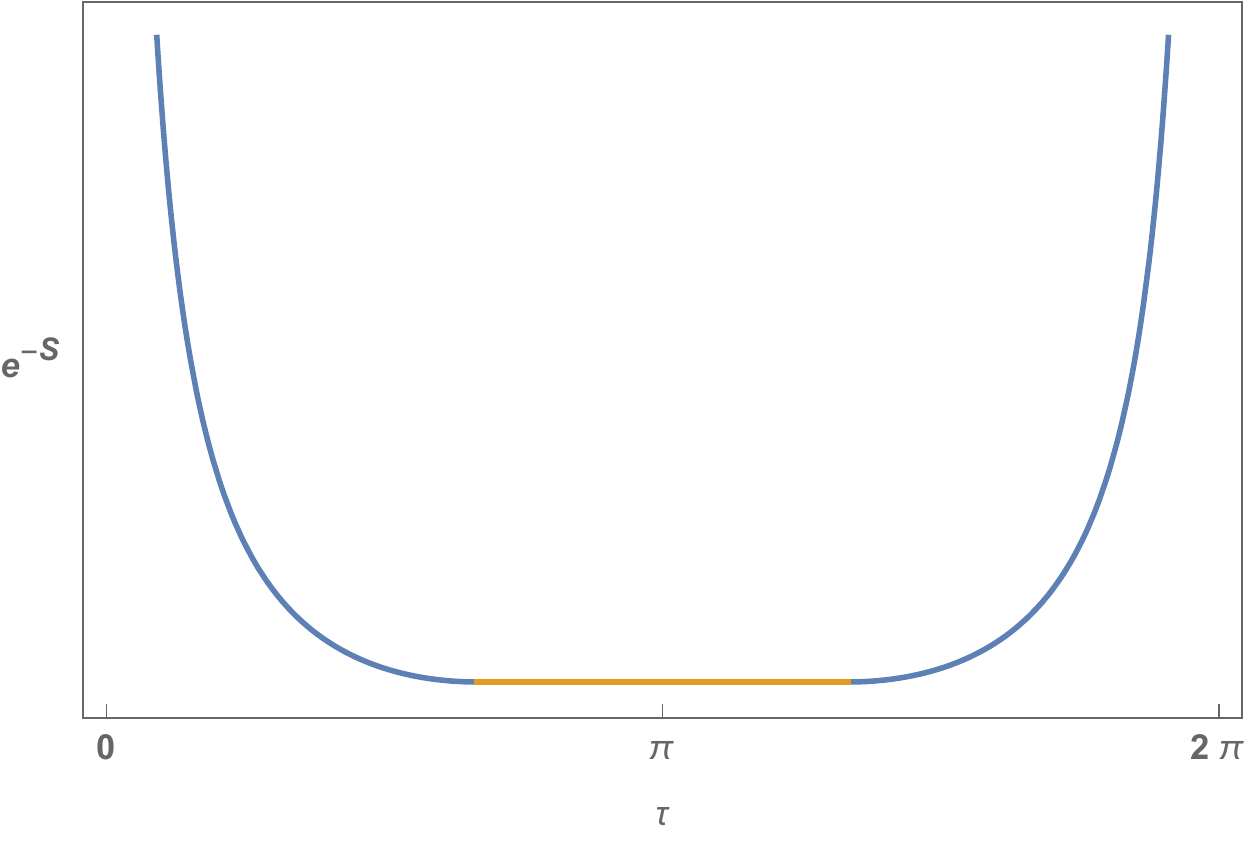}
      \caption{Correlator in 2d (we strip off numerical factors). In blue/yellow we depict the region where the U-shape/straight configuration dominates.}
      \label{correlator2d}
\end{figure}

If we consider the more general case of generic $\Delta_k$, then \eqref{mu} gets modified by changing the $\pm \frac{1}{2}$ factor into $\pm \frac{\Delta_k}{2\Delta}$. As a consequence, the central interval where the straight configuration dominates is of length 

\begin{equation}
\Delta \tau_{\rm straight}=\pi-\arccos\Big(\frac{\Delta_k}{2\Delta}\Big)\,.
\end{equation}
In the limit $\Delta\gg \Delta_k$ this length goes to 0, so the U-shaped configuration is valid everywhere --at $\tau=\pi$ it becomes degenerate with the straight one-- and we recover the case previously studied in \eqref{Dk<<D}.

\subsection{Time dependent correlators for $\Delta_i=\Delta_j$ in $d=4$}

Let us now consider $d=4$. For simplicity, we will consider diagonal correlators $\Delta_i=\Delta_j=\Delta$. This again corresponds to a symmetric configuration where $\mu\equiv\mu_i=-\mu_j$, and

\begin{equation}
    z_I=\sqrt{\frac{-2\Delta^2\mu^2+\sqrt{(4\Delta^2-\Delta_k^2)^2+4\Delta^4\mu^4}}{4\Delta^2-\Delta_k^2}}\,.
\end{equation}
Inserting this into the solution to the equations of motion one finds that the intersection point is

\begin{equation}
    \tau_I=\tau_i-\frac{1}{4}\log R+\frac{i}{4}\log I\,;\qquad \tau_I=\tau_j+\frac{1}{4}\log R-\frac{i}{4}\log I\,;
\end{equation}
where (we quote for simplicity the case $\Delta_k=\Delta$)

\begin{equation}
R=\frac{(2+\mu(2+\mu))(3-2\mu^2+\sqrt{9+4\mu^2})}{2\mu^4-\mu^2(1+\sqrt{9+4\mu^2})+2(3+\sqrt{9+4\mu^2})+2\mu|\mu|\sqrt{-2\mu^2+\sqrt{9+4\mu^2}}}\,;
\end{equation}
and

\begin{equation}
    I=\frac{(-2+\mu(-2i+\mu))\,(-3-2\mu^2+\sqrt{9+4\mu^2})}{-2(3+\sqrt{9+4\mu^2})+\mu\big( \mu-2\mu^3+\mu \sqrt{9+4\mu^2}+2|\mu|\sqrt{-2\mu^2+\sqrt{9+4\mu^2}}\big)}\,.
\end{equation}
Demanding that the geodesic arcs meet leads to ($\tau=\tau_i-\tau_j$)

\begin{equation}
\label{eq4d}
    \tau=\frac{1}{2}\log R-\frac{i}{2}\log I\,;
\end{equation}
One can now check that $|I|=1$, so $I=e^{i\theta}$. Since $\theta$ is defined modulo $2\pi$, it then follows that $\tau$ is defined modulo $\pi$. Restoring the factors of temperature, this implies the equivalence under $\tau\rightarrow \tau+\beta$, which is the expected KMS periodicity.

Let us now explore the small $\tau$ region, corresponding to large $\mu$. Solving perturbatively (the general version of) eq.\eqref{eq4d}, one finds

\begin{equation}
    \mu^{-1}\sim \frac{\Delta}{2\Delta-\Delta_k}\,\tau+\frac{\Delta\,(32\Delta^2+18\Delta\Delta_k+3\Delta_k^2)}{160\,(2\Delta-\Delta_k)^3}\,\tau^5+\cdots\,.
\end{equation}
Plugging this into the action, we find

\begin{equation}
    \langle O_i(\tau)\,O_i(0)\rangle_{SL}\sim \langle O_k\rangle \lambda_{iik}\,\frac{e^{\pi^4\,\frac{32\Delta^2+18\Delta\Delta_k+3\Delta_k^2}{640\,(2\Delta-\Delta_k)}\,(T\tau)^4+\cdots}}{\tau^{2\Delta-\Delta_k}}\,.    
\end{equation}

Let us now discuss the correlator in the full range $\tau\in[0,\beta]$. Choosing first $\Delta_k=\Delta$, one can see from eq.\eqref{eq4d} that $\mu=0$ is attained for $\tau=\frac{\pi}{3},\,\frac{2\pi}{3}$ (which, recovering dimensions, translates into $\frac{\beta}{3},\,\frac{2\beta}{3}$). Just as in the 2d toy model, in the intermediate regime there is no U-shaped configuration, and thus $\tau\in[\frac{\beta}{3},\frac{2\beta}{3}]$ is described by the straight configuration. Reducing the ratio $\frac{\Delta_k}{\Delta}$ reduces the size of this interval where the straight configuration dominates, and in the limit $\Delta_k\ll \Delta$ the interval is only the point $\tau=\frac{\beta}{2}$ --where the straight and U-shaped configurations become degenerate. Moreover, consistently, one recovers eq.\eqref{Dk<<D} in this limit.

\subsubsection{Including the spatial dependence}\label{spatial}

We now want to consider correlators with the full spacetime dependence. We will again restrict for simplicity to the case $\Delta_i=\Delta_j=\Delta$. The saddle is given by the solution to eqs.\eqref{saddleeq} now with both $\mu_i,\nu_i$ being non-zero. This means that our geodesic will bend both in $\tau$ and in $\vec{x}$ (which, using $SO(3)$, we can align with say $x_1$). Moreover, since $\Delta_i=\Delta_j$ ensures that the geodesic will be symmetric, we can consider the operator insertions --\textit{i.e.} the boundary points from where the geodesic arcs depart-- at $(-\tau_1,-x_1)$, $(\tau_1,x_1)$, so that the intersection point is at $\tau=0,x=0$. Through the equations of motion, this provides the conditions which allow to relate, $(\mu,\nu)=(\mu_1,\nu_1)=(-\mu_2,-\nu_2)$ with $(\tau_1,x_1)$. 

We will be interested in the regime $(\tau_1,x_1)\ll \beta$, where the correlator is governed by the OPE. This corresponds to large $(\mu,\nu)$. Since the regime is $(T\tau_1,Tx_1)\ll 1$ we can regard $\tau_1$, $x_1$ fixed and explore this regime by solving the system in perturbation theory in $T$. To that matter we write

\begin{equation}
    \mu=\sum_{n=1} a_n\, T^n\,,\qquad \nu=\sum_{n=1}b_n\,T^n\,,\qquad z_I=\sum_nc_n\,T^n\,.
\end{equation}
Plugging this ansatz into the eoms. and saddle equation, and upon expansion in $T$, we can solve order by order algebraically for the coefficients $(a_n,b_n,c_n)$. Note that we have 2 constraints coming from the equations of motion (one for the meeting point in $\tau$ and another for the meeting point in $x$) and one saddle equation, which allows to fix the 3 coefficients at each order. Even though the intermediate expressions are very lengthy and unilluminating (and hence we will refrain from showing them), the procedure is completely straightforward. Evaluating then the action on the solution one finds

\begin{equation}
    \langle O_i(\tau,\vec{x})\,O_i(0)\rangle_{SL}\sim \langle O_k\rangle \lambda_{iik}\,\frac{e^{\pi^4\,\frac{32\Delta^2+18\Delta\Delta_k+3\Delta_k^2}{640\,(2\Delta-\Delta_k)}\,(T|x|)^4\,C_2^{(1)}(\eta)+\cdots}}{\tau^{2\Delta-\Delta_k}}\,;    
\end{equation}
where $|x|$ and $\eta$ are defined below eq.\eqref{leadingterm}. As anticipated above, we recover the $C_2^{(\frac{d}{2}-1)}$ Gegenbauer polynomial, which suggests that the subleading term captures the contributions of $T^nO_k$ to the OPE.

\section*{Acknowledgements}

We are thank Jorge Russo for comments on the draft, as well as for many useful conversations. We would like to thank Gabriel Cuomo, Martin Rocek for useful conversations. D.R-G is partially supported by the Spanish government grant MINECO-16-FPA2015-63667-P. He also acknowledges support from the Principado de Asturias through the grant FC-GRUPIN-IDI/2018/000174.

\begin{appendix}

\section{Quick reminder of geodesics in the black brane background}\label{appendix:geodesics}

In this appendix we review the relevant properties of geodesic arcs following \cite{Rodriguez-Gomez:2021mkk} with the slight generalization to consider geodesics not necessarily with one endpoint anchored to the boundary.

Consider a massive particle in the black brane background \eqref{blackbrane} travelling from point $x_1$ to point $x_2$. The (euclidean) action is\footnote{In these conventions, the WKB solution to the lorentzian propagator $e^{-iS}$, with $S$ the action for a particle of mass $\Delta$.}

\begin{equation}
\label{Sparticle}
S=-i\Delta\,\int dz\,\frac{1}{z}\,\sqrt{f(z)\,\dot{\tau}^2+\dot{\vec{x}}^2+\frac{1}{f(z)}}\,,
\end{equation}
where dot stands for $z$-derivative, and we have used $\Delta=mR$ for large $\Delta$. To begin with, with no loss of generality we can consider our geodesic arc to be aligned along the $x_1$ direction (denoted simply by $x$) in space (when joining various arcs, each will be aligned along the corresponding direction, but for each arc, we can just choose it to be along $x_1$). Since the action does not depend on $\tau$ nor on $\vec{x}$, their canonically conserved momenta (denoted by $P_{\tau}=i\,\Delta\, \mu$ and $P_x=i\,\Delta\,\nu$ \footnote{The $i$'s are simply due to the fact that we are including an extra $i$ in the euclidean action \eqref{Sparticle}.}) are conserved. This gives two first order equations

\begin{equation}
\label{eomstaux}
\dot{\tau}=\frac{z\,\mu}{f(z)\,\sqrt{f(z)-z^2\,\mu^2-f(z)\,z^2\,\nu^2}}\,,\qquad \dot{x}=\frac{z\,\nu}{\sqrt{f(z)-z^2\,\mu^2-f(z)\,z^2\,\nu^2}}\,.
\end{equation}
These are to be integrated with the boundary conditions that the geodesic passes through $x_1$ and $x_2$. It is clear that the point $z_{\rm max}$ given by

\begin{equation}
\label{zmax}
f(z_{\rm max})-z_{\rm max}^2\,\mu^2-f(z_{\rm max})\,z_{\rm max}^2\,\nu^2=0\,,
\end{equation}
corresponds to the maximal reach of a geodesic: the turning point of a would-be $U$ shaped geodesic.  This signals that there are actually two types of geodesics \cite{Rodriguez-Gomez:2021mkk}: in-going (going directly from one point to the other) and returning (departing the point closest to the boundary all the way to $z_{\rm max}$ and then coming back to the other point). In the case at hand we expect, nevertheless, that the relevant geodesics are in-going.

\subsection{An equivalent formulation}

For completeness, let us briefly review a perhaps more standard treatment of geodesics in the literature (see \textit{e.g.} \cite{Fidkowski:2003nf,Festuccia:2005pi,Hubeny:2006yu,Hubeny:2012ry}). To begin with, let us strip a $-i$ factor, so that the geodesic contributes with $e^{-\widetilde{S}}$. Moreover, let us describe the trajectory of the particle with a worldline parameter $s$, and re-write eq.\eqref{Sparticle} by introducing a worldline metric $e$ as

\begin{equation}
\label{Sparticle2}
\widetilde{S}=\Delta\,\int dz\,\sqrt{e}\,\Big(\frac{1}{2}e^{-1}\frac{1}{z^2}(f(z)\,\dot{\tau}^2+\dot{\vec{x}}^2+\frac{\dot{z}^2}{f(z)})+\frac{\Delta^2}{2}\Big)\,,
\end{equation}
where the dot stands for $s$-derivative. Upon integrating out $e$ one recovers eq.\eqref{Sparticle}. Instead, one may now gauge-fix $e=1$ and consider the action

\begin{equation}
\label{Sparticle3}
\widetilde{S}=\Delta\,\int dz\,\Big(\frac{1}{2}\frac{1}{z^2}(f(z)\,\dot{\tau}^2+\dot{\vec{x}}^2+\frac{\dot{z}^2}{f(z)})+\frac{\Delta^2}{2}\Big)\,,
\end{equation}
which has to be supplemented with the constraint arising from the $e$-eom

\begin{equation}
    \frac{1}{z^2}(f(z)\,\dot{\tau}^2+\dot{\vec{x}}^2+\frac{\dot{z}^2}{f(z)})-\Delta^2=0\,.
\end{equation}
From \eqref{Sparticle3} it is clear that the momenta conjugated to $\dot{\tau}$ --$P_{\tau}=\Delta\mu$-- and $\dot{\vec{x}}$ --$\vec{P}=\Delta\vec{\nu}$-- are conserved, so

\begin{equation}
\label{momenta}
    \mu=\frac{1}{z^2\,f(z)}\,\dot{\tau}\,,\qquad \vec{\nu}=\frac{1}{z}\,\dot{\vec{x}}\,.
\end{equation}
The constraint becomes

\begin{equation}
\label{constraint}
    \dot{z}^2=\Delta^2\,z^2\,f(z)\,\Big( 1-\frac{z^2\,\mu^2}{f(z)}-z^2\,\vec{\nu}^2\Big)\,.
\end{equation}
Using eq.\eqref{constraint}, eqs.\eqref{momenta} can be casted as equations for $z$-derivatives. When the dust settles, one ends up with \eqref{eomstaux}. Note as well that \eqref{constraint} shows that the turning points of the geodesic are precisely at \eqref{zmax}.

\subsection{Ingoing geodesics}

Consider an ingoing geodesic departing $x_1=(\tau_1,x_1,z_1)$. Formally integrating the equations of motion gives

\begin{equation}
\tau=\tau_1+\int_{z_1}^z dz\frac{z\,\mu}{f(z)\,\sqrt{f(z)-z^2\,\mu^2-f(z)\,z^2\,\nu^2}}\,,\qquad x=x_1+\int_{z_1}^z dz\frac{z\,\nu}{\sqrt{f(z)-z^2\,\mu^2-f(z)\,z^2\,\nu^2}} \,.
\end{equation}
That the geodesic makes it to $x_2=(\tau_2,x_2,z_2)$ requires

\begin{equation}
\label{eoms}
\tau_2-\tau_1=I_{\tau}\,,\qquad x_2-x_1=I_x \,;
\end{equation}
with

\begin{equation}
I_{\tau}=\int_{z_1}^{z_2} dz\frac{z\,\mu}{f(z)\,\sqrt{f(z)-z^2\,\mu^2-f(z)\,z^2\,\nu^2}}\,,\qquad I_x=\int_{z_1}^{z_2} dz\frac{z\,\nu}{\sqrt{f(z)-z^2\,\mu^2-f(z)\,z^2\,\nu^2}}\,.
\end{equation}

Plugging the formal solution to the eom. in the action one finds

\begin{equation}
\label{G}
G=e^{-iS}=e^{-\Delta\,S_{\rm arc}}\,,
\end{equation}
with 

\begin{equation}
S_{\rm arc}=\int_{z_1}^{z_2}\,dz\frac{1}{z\,\sqrt{f(z)-z^2\,\mu^2-f(z)\,z^2\,\nu^2}}\,.
\end{equation}

This defines the geodesic implicitly through \eqref{eoms}. Note that these equations show that $\mu,\,\nu$ can be thought as functions of $\tau=\tau_2-\tau_1$, $x=x_2-x_1$ and both $z_{1,2}$. Taking the differential of the eoms. one finds

\begin{align}
&\frac{\partial I_{\tau}}{\partial \mu}\,\frac{\partial\mu}{\partial\tau}+\frac{\partial I_{\tau}}{\partial \nu}\,\frac{\partial\nu}{\partial\tau}=1\,; &\frac{\partial I_{x}}{\partial \mu}\,\frac{\partial\mu}{\partial \tau}+\frac{\partial I_{x}}{\partial \nu}\,\frac{\partial\nu}{\partial \tau}=0\,; \\ & \frac{\partial I_{\tau}}{\partial \mu}\,\frac{\partial\mu}{\partial x}+\frac{\partial I_{\tau}}{\partial \nu}\,\frac{\partial\nu}{\partial x}=0\,; &  \frac{\partial I_{x}}{\partial \mu}\,\frac{\partial\mu}{\partial x}+\frac{\partial I_{x}}{\partial \nu}\,\frac{\partial\nu}{\partial x}=1\,; \\ & \frac{\partial I_{\tau}}{\partial z_{1}}+ \frac{\partial I_{\tau}}{\partial \mu}\,\frac{\partial\mu}{\partial z_{1}}+ \frac{\partial I_{\tau}}{\partial \nu}\,\frac{\partial\nu}{\partial z_{1}}=0\,; & \frac{\partial I_{x}}{\partial z_{1}}+ \frac{\partial I_{x}}{\partial \mu}\,\frac{\partial\mu}{\partial z_{1}}+ \frac{\partial I_{x}}{\partial \nu}\,\frac{\partial\nu}{\partial z_{1}}=0\,; \\ & \frac{\partial I_{\tau}}{\partial z_{2}}+ \frac{\partial I_{\tau}}{\partial \mu}\,\frac{\partial\mu}{\partial z_{2}}+ \frac{\partial I_{\tau}}{\partial \nu}\,\frac{\partial\nu}{\partial z_{2}}=0\,; & \frac{\partial I_{x}}{\partial z_{2}}+ \frac{\partial I_{x}}{\partial \mu}\,\frac{\partial\mu}{\partial z_{2}}+ \frac{\partial I_{x}}{\partial \nu}\,\frac{\partial\nu}{\partial z_{2}}=0\,.
\end{align}

This allows to evaluate the contribution of a given arc to the saddle point equations. For the $\tau,\,x$ variation (the derivatives with respect to $\tau_{1,2}$, $x_{1,2}$ follow in an obvious way)

\begin{equation}
 \frac{dS_{\rm arc}}{d\tau}=\frac{\partial S_{\rm arc}}{\partial \mu}\,\frac{\partial \mu}{\partial\tau}+\frac{\partial S_{\rm arc}}{\partial \nu}\,\frac{\partial \nu}{\partial\tau}\,; \qquad \frac{dS_{\rm arc}}{dx}=\frac{\partial S_{\rm arc}}{\partial \mu}\,\frac{\partial \mu}{\partial x}+\frac{\partial S_{\rm arc}}{\partial \nu}\,\frac{\partial \nu}{\partial x}\,.
\end{equation}
When the dust settles, using here the previous formulas yields

\begin{equation}
\frac{dS_{\rm arc}}{d\tau}= \mu \,;\qquad  \frac{dS_{\rm arc}}{dx}=\nu\,.
\end{equation}

In turn, a similar, computation for the variation with respect to $z_{1,2}$ yields

\begin{equation}
\frac{dS_{\rm arc}}{dz_{1,2}}=\epsilon \frac{1}{f(z)\,z}\,\sqrt{f(z)-z^2\,\mu^2-f(z)\,z^2\,\nu^2}\,;
\end{equation}
where $\epsilon=+1$ for $z_2$ and $\epsilon=-1$ for $z_1$.

\subsubsection{Non-space-dependent geodesics}

Consider the special case where $\nu=0$. Through the equation of motion \eqref{eoms}, this corresponds to a geodesic at a fixed value of $x$. In this case \eqref{zmax} reads explicitly

\begin{equation}
1-\frac{z_{\rm max}^d}{z_0^d}=z_{\rm max}^2\,\mu^2\,,
\end{equation}
which implies that $z_{\rm max}\leq z_0$, that is, the geodesic stays always outside the horizon. In the limiting case $\mu=0$ --which corresponds to a purely radial, ``straight" geodesic at fixed $\tau$-- the geodesic goes all the way to the horizon (which is the end of the space in the euclidean). Note that one may then imagine a configuration departing the boundary at an arbitrary $\tau$ and coming back to the boundary at another arbitrary $\tau$ by joining two of these straight arcs, each departing at the corresponding fixed and arbitrary $\tau$ at the boundary. Of course, this configuration has a ``wedge" --simply because the two straight geodesic arcs would not be aligned in $\tau$-- unless the two boundary $\tau$'s are oppositely located in the thermal circle. Thus this configuration is not acceptable as a contribution to the leading part of the 2-point function (unless the $\tau$ difference is half of the thermal circle, when it is degenerated with the honest U-shaped geodesic). However, our subleading contribution to the correlator comes from evaluating a Witten diagram with geodesic bits and does not necessarily require the final configuration to be a smooth geodesic.

\end{appendix}

\end{document}